\documentclass[lettersize,journal]{IEEEtran}
\usepackage{graphicx}
\usepackage{subfigure}
\usepackage{float}
\usepackage{amsmath}
\usepackage{epstopdf}
\usepackage{cases}
\usepackage{color}
\usepackage{algorithm}
\usepackage{algpseudocode}
\usepackage{amsmath}
\usepackage{amssymb}
\usepackage{caption}
\makeatletter
\renewcommand{\maketag@@@}[1]{\hbox{\m@th\normalsize\normalfont#1}}%
\makeatother
\usepackage{stfloats}
\usepackage{cite}
\usepackage{makecell}
\usepackage{multirow}
\usepackage{ulem}

\ifCLASSINFOpdf
\else
\fi

\usepackage{caption}
\usepackage{mathtools}

\begin{document}
\title{Deep Cooperation in ISAC System: Resource, Node and Infrastructure Perspectives}
\author{
Zhiqing Wei,~\IEEEmembership{Member,~IEEE,}
Haotian Liu,~\IEEEmembership{Student Member,~IEEE,}
Zhiyong Feng,~\IEEEmembership{Senior Member,~IEEE,}\\
Huici Wu,~\IEEEmembership{Member,~IEEE,}
Fan Liu,~\IEEEmembership{Senior Member,~IEEE,}
Qixun Zhang,~\IEEEmembership{Member,~IEEE,}
Yucong Du,~\IEEEmembership{Student Member,~IEEE}\\

\thanks{
% This work is supported by This work was supported in part by the 
% National Natural Science Foundation of China (NSFC) under Grant 62271081, 
% in part by the National
% Key Research and Development Program under Grant 2020YFA0711302,
% in part by the 
% National Natural Science Foundation of China (NSFC) under Grant
% 92267202 and Grant U21B2014.		
This work was supported in part by the National Natural Science Foundation of China (NSFC) under Grant 62271081, U21B2014, and 92267202, and in part by the National Key Research and Development Program of China under Grant 2020YFA0711302. Title: ``Deep Cooperation in ISAC System: Resource, Node and Infrastructure Perspectives"
General Open Call for Articles. Vertical Area Editor: Khabbaz, Maurice. Guest Editor: Rabie, Khaled M.	\textit{Corresponding authors: Haotian Liu; Zhiqing Wei.}	

Zhiqing Wei, Haotian Liu,  Zhiyong Feng, Huici Wu, Qixun Zhang, and Yucong, Du
are with Beijing University of Posts and Telecommunications,
Beijing 100876, China (email: 
\{weizhiqing, haotian\_liu, fengzy, dailywu, zhangqixun, duyc\}@bupt.edu.cn). 

Fan Liu is with 
Southern University of Science and Technology, Shenzhen 518055, China
(e-mail: liuf6@sustech.edu.cn).
}}

\maketitle

\begin{abstract}
With the emerging Integrated Sensing and Communication (ISAC) technique, 
exploiting the mobile communication system with multi-domain resources,
multiple network elements, and large-scale infrastructures to 
realize cooperative sensing is a crucial approach satisfying the requirements of 
high-accuracy and large-scale sensing in Internet of Everything (IoE).
In this article, the deep cooperation in ISAC system 
including three perspectives is investigated.
In the microscopic perspective, namely, within a single node, the sensing information carried by time-frequency-space-code domain resources is processed, such as phase compensation, coherent accumulation and other operations, thereby improving the sensing accuracy.
In the mesoscopic perspective, the sensing accuracy could be improved through 
the cooperation of multiple nodes. We explore various multi-node cooperative sensing scenarios and present the corresponding challenges and future research trends. 
In the macroscopic perspective, 
the massive number of infrastructures from 
the same operator or different operators could perform cooperative sensing 
to extend the sensing coverage and improve the sensing continuity. 
We investigate network architecture, target tracking methods, and the large-scale sensing assisted digital twin construction.
Simulation results demonstrate the superiority of multi-nodes and multi-resources cooperative sensing over non-cooperative sensing.
This article may provide a deep and comprehensive 
view on the cooperative sensing 
in ISAC system to enhance the performance of sensing, 
supporting the applications of IoE.
\end{abstract}


\section{Introduction}

Internet of Things (IoT), Artificial Intelligence (AI), and automation technologies 
are reconfiguring traditional industries, 
% which are undergoing digital, networked, and intelligent transformation, 
opening the era of Internet of Everything (IoE).
The scenarios of IoE
are transferring from pure human to the symbiosis of human, intelligent machines, and massive 
number of sensors.
% , where 
% the information processing includes sensing, communication, computation, and
% decision-making, realizing the coupling of digital and physical spaces. 
These applications urgently need to be supported by new information infrastructures 
with the integration of sensing and communication. 
With the development of Integrated Sensing and Communication (ISAC) 
technique \cite{Liu_Survey},
the mobile communication system, 
as the crucial infrastructure to support the emerging IoE, 
is constantly breaking through the pure communication function
and integrating radar sensing function. 
Notably, the International Telecommunication Union (ITU) 
has identified ISAC as one of the scenarios of the Sixth-Generation (6G)
mobile communication system \cite{itu2022future}.

% motivation
The mobile communication system realizes radar sensing with ISAC technology 
by analyzing the echo of ISAC signal \cite{liu2020joint}, 
thereby realizing target localization, 
environment reconstruction, etc. 
ISAC not only enhances the utilization of spectrum and hardware resources, 
but also realizes the coupling of digital and physical spaces and 
the mutual benefit of communication and sensing functions \cite{Liu_Survey}. 
The Millimeter-wave (mmWave), Terahertz (THz), 
and massive Multiple Input Multiple Output (MIMO) technologies 
in mobile communication system are developing rapidly, 
which guarantee the feasibility of ISAC technique \cite{Wei_Survey}. 
However, in the applications of IoE, such as the sensing of 
vehicles and Unmanned Aerial Vehicles (UAVs), 
the fading and path occlusion of ISAC signals degrade the 
signal quality drastically, posing 
challenges to achieve high-accuracy, 
large-coverage, and continuous sensing.
To this end, it is urgent to explore cooperative sensing approaches 
within the ISAC system to enhance sensing performance.

Cooperative sensing in ISAC system could be realized from the following 
three comprehensive perspectives 
according to the scope of cooperation. 
\begin{enumerate}
    \item In the microscopic perspective, i.e., within a single node, 
    there still exists resource-level cooperation to improve sensing accuracy by fusing 
    the sensing information carried in the time-frequency-space-code multi-domain resources.
    \item In the mesoscopic perspective, 
     sensing accuracy could be improved through the cooperation of multiple nodes, including
     Base Station (BS), User Equipment (UE), UAV, and so forth. 
    \item In the macroscopic perspective, the large number of infrastructures from one operator 
    or even multiple operators could be applied to extend the sensing area 
    and improve the continuity of sensing.
\end{enumerate}

However, realizing the deep cooperation of ISAC system 
in the perspectives of resource, 
node and infrastructure, continues to face the following challenges.
\begin{enumerate}
    \item \textbf{Resource-level Cooperation:} 
    The inconsistency of physical-layer parameters on fragmented multi-domain resources 
    brings challenges to sensing information fusion. 
    For instance, differences in subcarrier spacings between high and low frequency bands hinder the seamless integration of sensing data collected from different frequency bands.
    \item \textbf{Node-level Cooperation:} 
    The fusion of the sensing information from multiple nodes 
    with non-synchronization or low-accuracy synchronization level
    in space, frequency and time domains is challenging.
    \item \textbf{Infrastructure-level Cooperation:} 
    The rapid and seamless handover among multiple BSs from 
    one operator or different operators is challenging.
\end{enumerate}

%related work and contribution
Facing the above challenges, there are some related studies. 
Zhang \textit{ et al.} in \cite{zhangPWN} initially 
introduced a concept of Perceptive Mobile Network (PMN) 
and proposed a Remote Radio Units (RRUs) cooperative sensing scheme 
under the architecture of Centralized Radio Access Network (C-RAN) 
to improve sensing accuracy. 
In \cite{Ji}, Ji \textit{et al.} proposed a broad concept of 
cooperative sensing in ISAC system, 
including multi-static, multi-band, and multi-source cooperation. 
Tong \textit{et al.} in \cite{Tong} further presented the concept, 
algorithm, and demonstration of multi-view cooperative sensing 
in ISAC system. 
\cite{Li, Wei} explored the sensing information fusion algorithms 
with multi-BS cooperative sensing. 
Zhang \textit{et al.} in \cite{zhang2024target} proposed multi-node cooperative passive sensing system and a sensing information fusion localization scheme.
Overall, existing studies on cooperative sensing in ISAC system 
mainly focus on multiple BSs cooperative sensing. 
However, even in terms of node-level cooperative sensing, 
such as our previous work \cite{Wei}, 
the cooperation between BS and UE, the cooperation between macro BS and micro BS, and the cooperation between BS and UAV 
are rarely studied.
Besides, the multi-BS cooperation is not structurally classified. 

This article aims to provide 
a deep, comprehensive and concise view on cooperative sensing in ISAC system, 
namely, the cooperation in the perspectives of 
resource, node and infrastructure. 
The main contributions of this article are as follows.

\begin{enumerate}
    \item In terms of resource-level cooperation, 
    the echo signals in multiple antennas, 
    multiple time slots or frames,
    multiple frequency bands, and multiple code words could be fused to improve sensing accuracy.
    Meanwhile, the sensing information in multi-domain resources could be fused simultaneously to 
    further enhance the sensing accuracy.
    
    \item In terms of node-level cooperation, 
    we have provided a concise classification and 
    structural relation for multi-node cooperative sensing.
    Multiple macro BSs or multiple micro BSs could perform cooperative sensing.
    Given the proximity to the target and potential deployment on high-frequency bands, micro BSs exhibit higher sensing accuracy compared to macro BSs.
    Hence, multiple micro BSs cooperative sensing is a preferred scheme.
    Nevertheless, multiple micro BSs have smaller overlapping areas. When the target moves beyond the overlapped coverage of multiple micro BSs, cooperation between macro and micro BSs becomes necessary.
    Meanwhile, the cooperation schemes between BS and UE
    are also studied in this article.
    Multi-node cooperation effectively improves sensing accuracy and continuity.

    \item In terms of infrastructure-level cooperative sensing,
    the infrastructures from one or multiple operators, 
    across the air, ground and space,
    could be applied to realize seamless target sensing.
    The network architecture supporting cooperative sensing, 
    the moving target detection and tracking methods, 
    and the large-scale sensing assisted digital twin construction, 
    are investigated in the infrastructure-level cooperation.
\end{enumerate}

The remainder of this article is organized as follows. Section \ref{se2} outlines the basic concepts in ISAC system. Sections \ref{se3}, \ref{se4}, and \ref{se5} details the resource-level, node-level, and infrastructure-level cooperation, respectively. Section \ref{se6} provides the performance evaluation. And Section \ref{se7} concludes the article by summarizing the key viewpoints.

\begin{figure*}[!ht]
    \centering
    \includegraphics[width=0.85\linewidth]{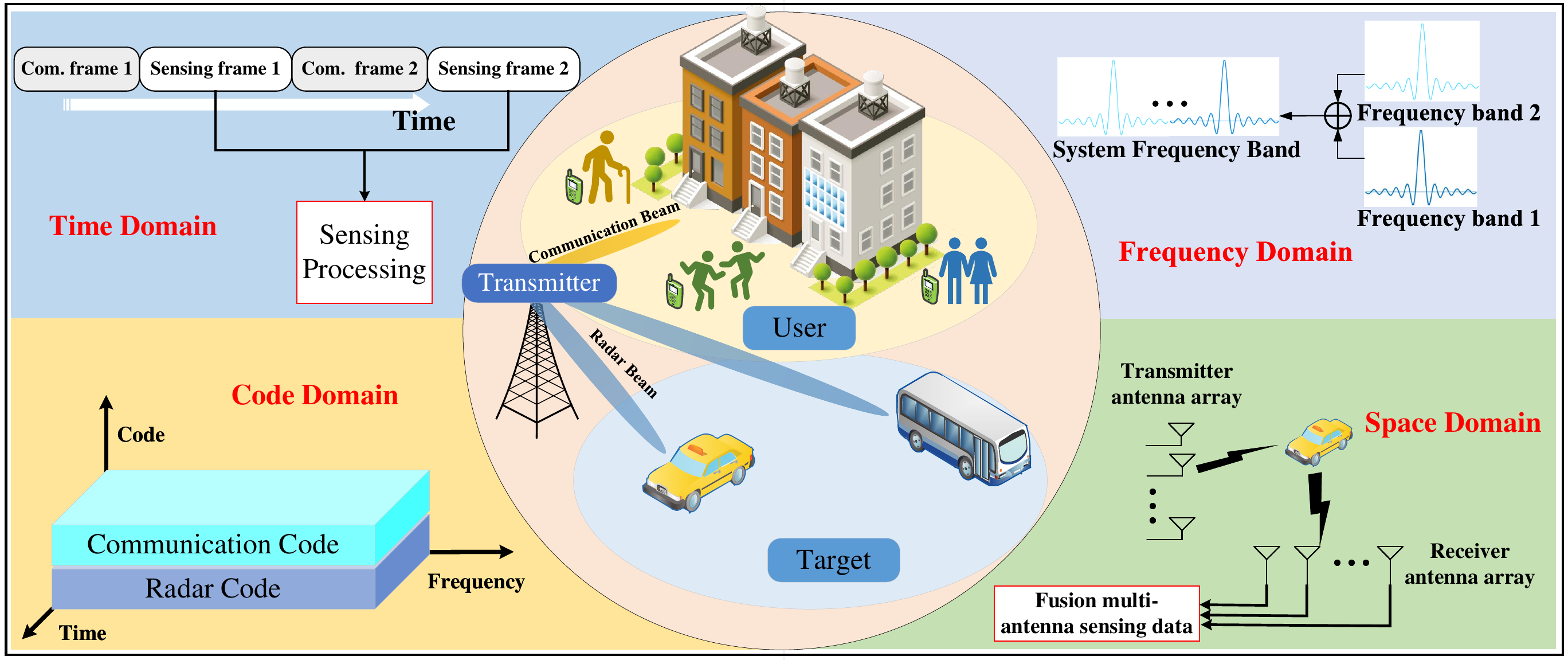}
    \caption{Resource-level cooperation in multi-domain resources.}
    \label{fig.1}
\end{figure*}

\section{Basic Concepts in ISAC System}\label{se2}
In this section, we provide a brief overview of the types of sensing, performance metrics, scenarios, and requirements within the ISAC system.

\subsection{Types of Sensing}
The radar sensing in ISAC system mainly includes 
active sensing and passive sensing. 
In active sensing, the transceiver detects the target by 
receiving the echo signal reflected by the target. 
In passive sensing, the receiver detects the target by receiving 
the signal emitted from the target or the echo signal from other transmitters 
reflected by the target \cite{zhangPWN}.
The BS with sensing function realizes the coupling of 
digital space and physical space, 
which is a unified information infrastructure supporting IoE.

\subsection{Performance Metrics of Sensing} \label{sec2-B}
\begin{itemize}
    \item \textbf{Accuracy of detection:} 
    The accuracy of detection is measured by 
    the probabilities of detection and false alarm. 
    This performance metric is primarily applied in the scenario of intrusion detection for UAVs and pedestrians.
    \item \textbf{Accuracy of parameter estimation:} 
    The accuracy of the estimation of distance and velocity of target 
    is measured by Mean Square Error (MSE), Rooted MSE (RMSE) 
    and Normalized MSE (NMSE), 
    which measure the deviation between 
    the estimation value and the real value.
    This performance metric is used to evaluate the feasibility of obstacle avoidance for vehicles and environmental reconstruction. It is also used to evaluate the effectiveness and superiority of sensing signal processing methods.
    \item \textbf{Sensing area:} 
    Sensing area measures the range of radar sensing, 
    which is influenced by the fading of radar signal and 
    the Radar Cross-Section (RCS) of target.
    The assessment of sensing area is a prerequisite for the operation of sensing service, which simultaneously facilitates the implementation of multi-node cooperation.
    \item \textbf{Sensing continuity:} 
    Sensing continuity measures 
    the performance of target tracking. 
    When detecting the target in $n$ continuous time instants, 
    the fraction of the time instants with sensing accuracy 
    higher than a threshold to $n$ is defined 
    as the sensing continuity. 
    Sensing continuity is a performance metric applied in target tracking scenarios.
\end{itemize}

\subsection{Scenarios and Requirements}
\begin{itemize}
    \item \textbf{Cooperative Sensing of Vehicles:} 
    In the application of intelligent transportation, 
    the lane-level sensing of vehicles needs to be realized. 
    Thus, cooperative sensing is essential in this scenario to improve 
    the sensing accuracy.
    \item \textbf{Cooperative Sensing of UAVs:} 
    In the scenario of UAV sensing in the urban areas, 
    the cooperative sensing is required to realize continuous sensing of 
    UAVs with high maneuverability due to the blockage of buildings.
    Since the RCS of UAV is small, 
    the cooperative sensing is required to improve the accuracy of 
    small target sensing.
    \item \textbf{Cooperative sensing of other devices:}
    In the scenario of smart home and smart industry, high-accuracy, comprehensive, and low-latency sensing services are required for the services such as human respiratory detection, illegal intrusion detection, environmental reconstruction in factories, etc. Therefore, cooperative sensing among Wi-Fi, indoor small cells, terminals, and other sensing devices is essential.
\end{itemize}

\section{Resource-level Cooperation}\label{se3}

As shown in Fig. \ref{fig.1}, 
within a single node, the resource-level cooperation 
could be realized to improve the accuracy, resolution of sensing, which improves the accuracy of detection and parameter estimation in Section~\ref{sec2-B}.
Resource-level cooperation, as the cooperation in microscopic perspective, could be combined with the other levels of cooperation to further improve the sensing performance. 

\subsection{Cooperation in Space-domain Resource}\label{sec3.1}

Resource cooperation in space-domain mainly apply MIMO 
to exploit the multi-path signal propagation. 
As shown in Fig. \ref{fig.1}, the signals in multiple antennas could be fused 
to enhance the accuracy of sensing, which is further classified into the sensing information fusion 
with uniform and 
non-uniform antenna array. 
Meanwhile, multi-antenna beamforming can reduce the interference between communication and sensing.

\subsubsection{Sensing information fusion over uniform antenna array} 

In single-node sensing scenario, the angle, 
distance and velocity of target need to 
be estimated for localization and trajectory prediction. 
With a uniform antenna array, the number of antennas affects the resolution of angle estimation for the target, 
which is realized using Multiple Signal Classification (MUSIC) method, 
Estimating Signal Parameter via Rotational Invariance Techniques (ESPRIT) method, 
etc \cite{Wei_Survey}. 
On the other hand, the accuracy and resolution of 
distance estimation are 
related to bandwidth and Signal-to-Noise Ratio (SNR) of echo signal, 
and the accuracy and resolution of velocity estimation 
are related to sensing time and SNR. 
Since multiple antennas occupy the same time-frequency resources, 
the SNR of echo signal can be improved 
by fusing the sensing information on multiple antennas,
further improving the performance of distance and velocity estimation,
which is realized by phase compensation and correlation accumulation. 
Meanwhile, the time-frequency resources are 
extended by utilizing the data of multi-antenna, 
i.e., enlarging the bandwidth and extending sensing time,
which is accomplished by splicing the data in 
each antenna with the challenge of 
determining the initial phase of the data in 
each antenna to perform splicing operation.

\subsubsection{Sensing information fusion over non-uniform antenna array} 

For non-uniform antenna arrays, 
such as sparse antenna arrays, 
optimization algorithms can be employed to achieve 
the performance of a full antenna array using only 
a portion of antenna resources.
For the estimation of target's angle, 
the virtual antenna array is obtained by 
hybrid cross-multiplication among antennas, 
so that the angle can be estimated using MUSIC method.
In contrast to uniform antennas, 
hybrid cross-multiplication leads to a decrease in the SNR on multi-antenna, 
which reduces the SNR gain in the sensing information fusion of 
multi-antenna and 
further degrades the sensing performance. 
Hence, the trade-off between improving sensing performance of 
multi-antenna fusion  
and saving space-domain resources exists with the non-uniform antenna array. 

\subsection{Cooperation in Time-domain Resource}

The cooperation in time-domain, as shown in Fig. \ref{fig.1}, 
is mainly used to improve the SNR of echo signal by time accumulation,
further improving the estimation accuracy of 
distance and velocity.
Assuming that ISAC system adopts 
Orthogonal Frequency Division Multiplexing (OFDM) signal, 
the multiple frames can be used for coherent 
or non-coherent accumulation 
to improve the SNR of echo signal and further improve the estimation 
accuracy of distance.
Similarly, 
multiple frames can be used to improve the resolution 
of velocity estimation.

\subsection{Cooperation in Frequency-domain Resource}\label{sec3.3}

\begin{figure*}[!ht]
	\centering
    \includegraphics[width=0.75\linewidth]{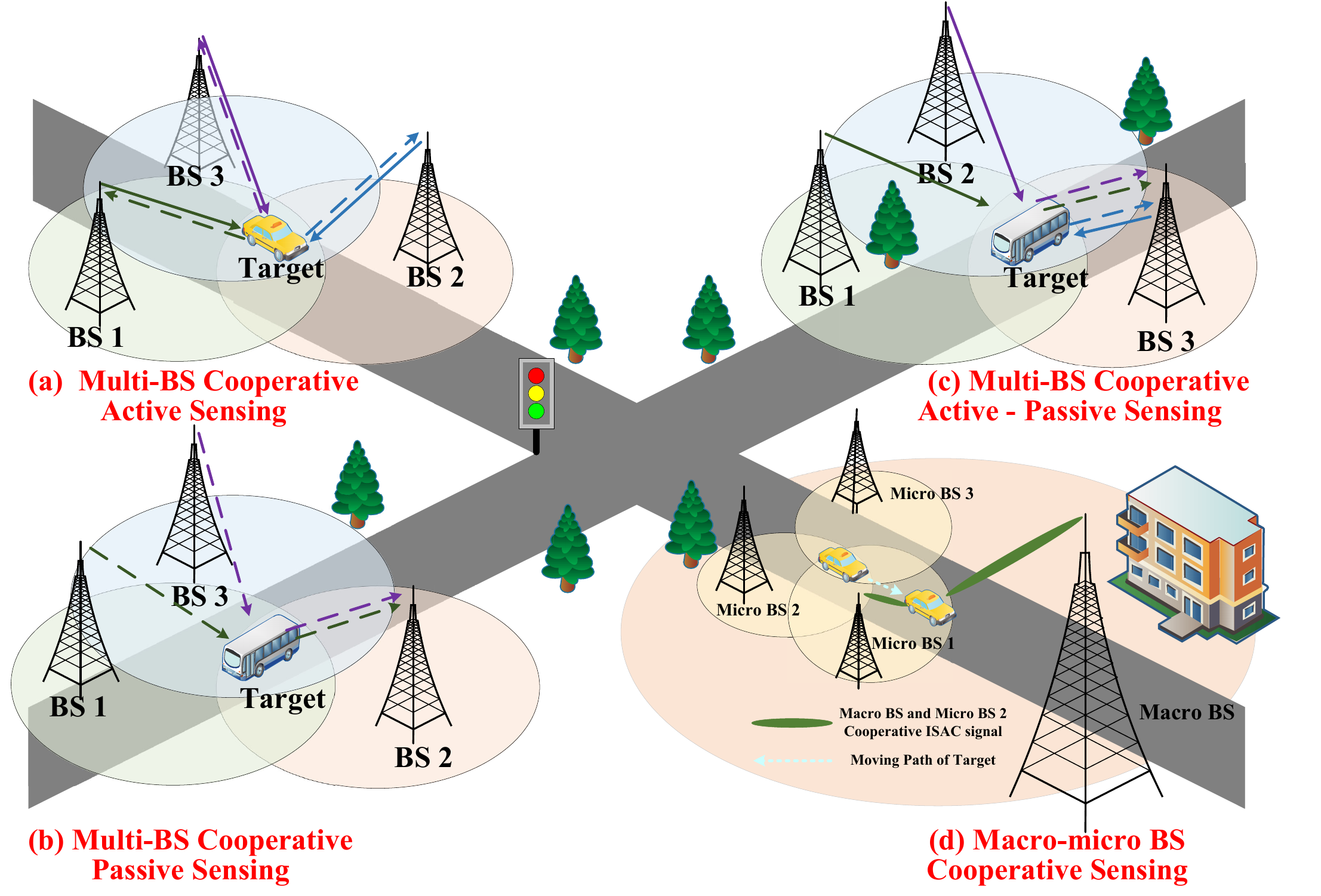}
    \caption{Multi-BS cooperation.}
    \label{fig.2}
\end{figure*}

Cooperation in frequency-domain mainly involves 
the fusion of the multiple reference signals and the signals on 
multiple frequency bands, 
as shown in Fig. \ref{fig.1}.

\subsubsection{Fusion over multiple reference signals}

Reference signals, also known as pilot signals, 
can be applied in radar sensing \cite{wei2023multiple}, 
which include 
downlink and uplink reference signals.
Downlink reference signals could be applied in 
downlink active or passive sensing.
The uplink reference signals including the 
Demodulation Reference Signal (DMRS), Sounding Reference Signal (SRS),
and so on, could be applied in uplink sensing. 
Since different reference signals occupy different resources 
in frequency domain, 
the various reference signals can be cooperatively 
used in radar sensing, 
achieving higher sensing performance than single reference signal.
Given that multiple reference signals do not fill 
the complete time-frequency resource blocks, 
the challenge of this research is the sidelobes deterioration 
caused by 
non-continuous resources in time-frequency domains.
The sparsity of targets when mapping the data in the 
time-frequency domain to the Doppler-delay domain brings 
an opportunity for the application of Compressed Sensing (CS) 
technique into ISAC signal processing. 
The CS technique can be applied to recover sensing information 
at empty position in the time-frequency resource blocks. 
Therefore, the CS-based multiple reference signals cooperative 
sensing has significant performance improvement.

\subsubsection{Fusion over high and low frequency bands}

The mmWave and THz frequency bands
are gradually applied in future mobile communication system. 
Meanwhile, the sub-6 GHz frequency bands are essential in enhancing the 
coverage of mobile communication system.
With the technique of Carrier Aggregation (CA), 
the high and low-frequency bands 
are combined to improve the performance of communication.
Similarly, the cooperation of high and low-frequency bands
can also improve the sensing performance, 
where the challenge is the inconsistency of physical layer parameters 
in the high and low-frequency bands during sensing information fusion. 
For example, when adopting OFDM as the ISAC signal, 
the different subcarrier spacings in the 
high and low-frequency bands brings challenge
for sensing information fusion.
This problem can be solved by reorganizing the 
channel information matrices of high and low-frequency bands 
to match the parameters of the corresponding positional elements \cite{WeiCA}.
In addition to the cooperation of high and low-frequency bands, 
there are additional cases of cooperation in frequency-domain, 
such as the fusion of the signals over the 
fragmented spectrum bands or unlicensed frequency bands.

\subsection{Cooperation in Code-domain Resource}

With the cooperation in code-domain, 
communication and radar sensing functions share the same resources 
in time-frequency domains, distinguished by code words. 
This approach enables higher resolution of distance and velocity 
estimation \cite{liu2023complementary},
as the resources in time-frequency domains are not diminishing
with the code-division method compared with the traditional 
time-division or frequency-division method.
Furthermore, the sensing information on different code words 
within the same time-frequency domains
can be extracted and fused to obtain high sensing accuracy.

\subsection{Cooperation in Multi-domain Resources}
In mobile communication systems, it is common to enhance communication performance through the utilization of multi-domain resources, which makes the multi-domain cooperative sensing feasible.
The cooperative sensing in multi-domain fuses the 
sensing information in multi-domain resources 
to achieve a higher sensing accuracy than the
sensing method using single-domain resource.
For example, using multiple antennas, 
multiple signals on different antennas may overlap 
in the time-frequency domains,
so that different code words are applied to distinguish 
the multiple signals.
Then, the multiple signals with different code words 
could be extracted 
and fused, thereby improving the sensing accuracy.

In urban environments, 
the sensing information in multi-path could be 
applied to improve the sensing accuracy 
with multi-domain resource cooperation.
The challenge of multi-path research is multi-path separation. Common methods include blind source separation and independent component analysis. However, the features of different paths are generally varied in multi-domain. Therefore, the multi-path mixed signals can be separated through multi-domain resource cooperation,
which are further fused to realize high-precision target localization, high-continuity target tracking, 
and high-accuracy environmental reconstruction.

\section{Node-level Cooperation}\label{se4}

\begin{figure*}[!ht]
	\centering
    \includegraphics[width=0.75\linewidth]{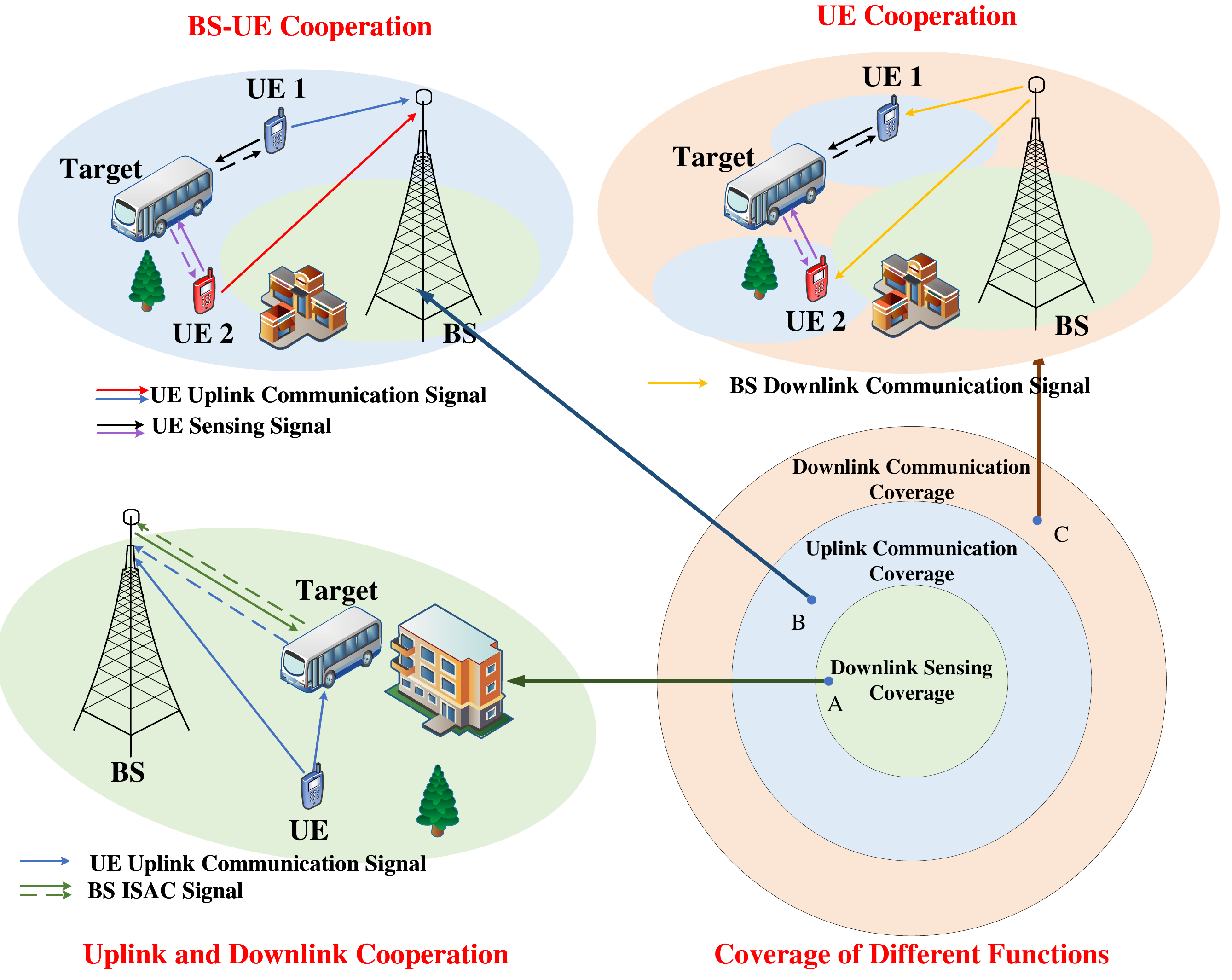}
    \caption{BS-UE cooperation.}
    \label{fig.3}
\end{figure*}

In node-level cooperation, multiple nodes, 
including macro BS, micro BS, UE, could cooperate to 
improve the accuracy of detection, the accuracy of parameter estimation, and the sensing area in Section~\ref{sec2-B}. Besides, node-level cooperation can be combined with resource-level cooperation to further improve the sensing performance. However, compared with resource-level cooperation, the node-level cooperation obtains the capability of multi-view sensing, which has potential in three-dimensional (3D) imaging for the target and environment. Therefore, compared to resource-level cooperation, node-level cooperation is a cooperation in mesoscopic perspective.

\subsection{Multi-BS Cooperation}
The cooperation between multiple BSs
is categorized according to the type of BS, 
namely multiple micro BSs cooperative sensing, 
macro-micro BSs cooperative sensing, 
and multiple macro BSs cooperative sensing,
where the sensing area is increasing in these three categories.

Fig. \ref{fig.2}(a) shows multi-BS cooperative active sensing, i.e., 
when the target is located in the overlapped coverage area of multiple BSs, 
each BS performs sensing separately, 
and the sensing information fusion is performed 
in the fusion center, which could be one BS or 
the Mobile Edge Computing Server (MECS) \cite{Wei_Networking}.
The challenges of cooperative active sensing lie in spatial registration and symbol-level fusion of multiple non-coherent echo sensing data.
Fig. \ref{fig.2}(b) shows multi-BS cooperative passive sensing, 
i.e., the passive BS receives the echo signals of other BSs 
reflected by the target 
and performs sensing information fusion. 
The challenges of cooperative passive sensing are how to eliminate the synchronization error and achieve a symbol-level fusion sensing.
Fig. \ref{fig.2}(c) shows 
multi-BS cooperative active-passive sensing, i.e., 
the BS can both perform active sensing and receive the echo signals 
of other BSs 
reflected by the target in passive sensing, 
and finally fuse the sensing information
of active sensing and passive sensing. 
The challenges of cooperative active sensing and cooperative passive sensing both exist in this type of cooperation.
The above three types of cooperative sensing 
requires that the target is located in the overlapped coverage 
area of multiple BSs.  
The future research trends in multi-BS cooperation include multi-BS mutual interference mitigation, multi-BS cooperative tracking, beamforming optimization, multi-BS sensing information fusion, and multi-BS space-time-frequency synchronization. 
In addition, since each BS can independently perform resource-level cooperation, the resource allocation in resource-level cooperation is also the future research trend.
Since the sensing area of micro BS is smaller than that of 
macro BS,
when the target moves out of the sensing area of micro BS,
the macro-micro BSs cooperative sensing is performed, as shown in 
Fig. \ref{fig.2}(d), 
which is investigated in Section \ref{sec4.2}.

\subsection{Macro-micro BSs Cooperation}\label{sec4.2}

Compared with macro BS,
micro BS has higher frequency band and smaller transmit power. 
Hence, the sensing area of 
micro BS is smaller than that of macro BS.
When the target moves out of the overlapped sensing area of multiple micro BSs
as shown in Fig. \ref{fig.2}(d), 
the macro BS and micro BS could cooperate to improve 
the sensing accuracy and continuity. The type of macro-micro BSs cooperation also consists of cooperative active, passive, and active-passive sensing. Macro-micro BSs cooperation not only encounters the challenges similar to those in multi-BS cooperation but also faces the challenge of
the fusion of the sensing information from the 
micro BS working on high-frequency band and the macro BS 
working on low frequency band.
To address the above challenge, 
the channel information matrices from macro and micro BSs are adjusted and
fused with low synchronization accuracy,
which achieves higher sensing accuracy 
compared with data-level sensing information fusion.

\subsection{BS-UE Cooperation}

The BS and UE cooperative sensing
can be classified according to the sensing range.
Generally, the downlink sensing range is smaller than 
the uplink communication range, 
and the uplink communication range
is smaller than the downlink communication range,
as shown in Fig. \ref{fig.3}.
When the target is located within the downlink sensing coverage, 
such as the location A in Fig. \ref{fig.3}, 
the BS can simultaneously perform downlink sensing by receiving the echo signal of BS
and uplink sensing by receiving the uplink reflected signal from UE.
Then, the echo signals from downlink and uplink sensing are fused in the BS,
which is the cooperative downlink and uplink sensing. 
When the target is outside of the downlink sensing coverage and 
within the uplink communication coverage, 
such as the location B in Fig. \ref{fig.3}, 
multiple UEs detect the target and upload the sensing information 
to the BS for sensing information fusion. 
When the target is outside of the uplink 
communication coverage and within
the downlink communication coverage, 
such as the location C in Fig. \ref{fig.3}, 
multiple UEs detect the target and fuse the sensing information by themselves,
with the guidance of BS in resource allocation.
The challenge of BS and UE cooperative sensing is to mitigate the uplink and downlink interference, achieving the mutual benefits of sensing and communication.
Hence, redesigning the frame structure is essential, coupled with leveraging uplink sensing information for beamforming.

\begin{figure*}[!ht]
	\centering
    \includegraphics[width=0.75\linewidth]{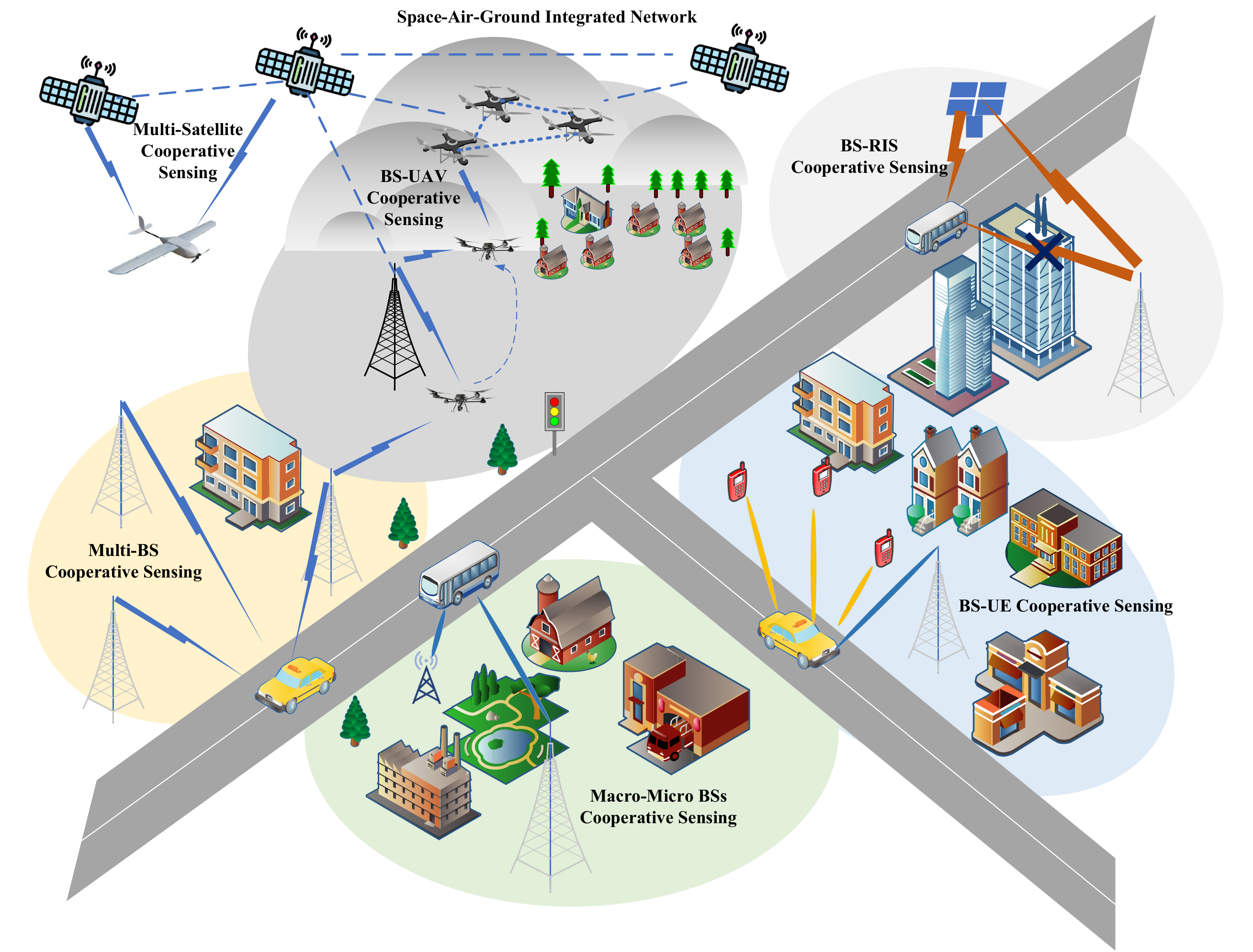}
    \caption{Infrastructure-level cooperation.}
    \label{fig.4}
\end{figure*}

\section{Infrastructure-level Cooperation}\label{se5}
In the application of smart city, 
there are a large number of targets to be detected and tracked. 
For example, in the scenario of UAV sensing, 
due to the high maneuverability of UAV, 
the infrastructure-level 
cooperation is required to accurately detect and track UAV. 
Infrastructure-level 
cooperation is expected to mainly enhance the sensing continuity in Section~\ref{sec2-B}, which achieves wider area sensing compared with the node-level and resource-level cooperation, representing the cooperation in macroscopic perspective.
Generally, the following techniques are 
required in the infrastructure-level cooperation.

\subsubsection{Network architecture supporting cooperative sensing}

In order to support ISAC enabled cooperative sensing, 
the network architecture needs to be designed. 
The network elements include Remote Radio Unit (RRU), 
Building Base Band Unite (BBU), MECS, and core network, etc. 
The sensing information fusion centers are deployed 
in the MECS for node-level cooperative sensing and 
deployed in the core network for infrastructure-level cooperative sensing. 
Multiple RRUs detect target with multi-domain resources and 
fuse the sensing information in the MECS. 
The interfaces between BSs, information processing procedures, and 
signaling interaction procedures 
among the network elements need to be designed to support cooperative sensing. 
The fusion center in the core network fuse 
the large amount of sensing information to 
build a full view of the digital twin of physical space.

In order to realize large-scale and seamless sensing, 
the Heterogeneous Networks (HetNets), 
including the BSs from different operators and 
the Access Points (APs) using IEEE 802 techniques,
could cooperate to sense target and environment. 
In this case, the network architecture supporting HetNets 
cooperative sensing needs to be designed.
A new framework for fusing sensing information from HetNets 
needs to be constructed to support
network interoperability.
Then, the procedure of sensing information fusion with 
non-synchronization among HetNets faces great challenges.
Hence, the space and time calibration among HetNets is necessary.

\subsubsection{Moving target detection and tracking} 

When tracking the target with high maneuverability, 
the infrastructures from the space-air-ground integrated networks, 
as well as the infrastructures from single or multiple operators, 
could cooperatively detect and track the target. 
As shown in Fig. \ref{fig.4}, multi-BS cooperation, 
as well as the cooperation 
between BS and UAV, could be performed to detect 
the vehicle and UAV with high maneuverability.
In this scenario, the handover of target sensing by 
multiple BSs needs to be studied.
The handover in target sensing 
studies the switching of multiple BSs belonging to the same operator 
or different operators to realize the continuous 
sensing of moving target
due to the limited coverage of single BS. 
Besides, the detection methods of moving small target, 
and the machine learning techniques 
for target tracking need to be studied. 

\subsubsection{Digital twin assisted infrastructure-level cooperation} 
In the scenario of intelligent transportation, 
the massive amount of sensing information 
needs to be structurally stored and processed.
Furthermore, heterogeneous infrastructures bring diverse protocols and processing methods for sensing information.
Therefore, the storage, management and fusion of sensing information are difficult in the infrastructure-level cooperation.
Digital twin can not only replace tedious and inefficient manual management of sensing information, 
achieving low-cost, visual, and intelligent management of sensing information,
but also enables optimal distributed resources scheduling.
Digital twin is a potential solution
for massive sensing information fusion in the infrastructure-level cooperation~\cite{wei2024integrated}.

In infrastructure-level cooperation,
the construction and updating of digital twin face some challenges.
On the one hand, 
the large number of devices bring great difficulties 
for the low-delay construction and updating of digital twin entities.
In this case, one of the viable solutions is to construct and update the digital twin entities
in a distributed manner.
On the other hand, given the high mobility of vehicles and UAVs, as well as the frequent switching of different devices,
ensuring the low-delay updating of the sensing information 
and the continuity of sensing is necessary.

\section{Performance Evaluation}\label{se6}

In this section, 
we provide the simulation results of resource-level and 
node-level cooperative sensing, 
verifying the advantages of cooperative sensing. 
Among the performance metrics mentioned in Section \ref{sec2-B}, 
the performance metric of RMSE is commonly 
used to characterize sensing accuracy.

\subsection{Resource-level Cooperation}

\begin{figure}[!ht]
	\centering
	\subfigure[Range estimation.] {\label{fig5.a}\includegraphics[width=0.4\textwidth]{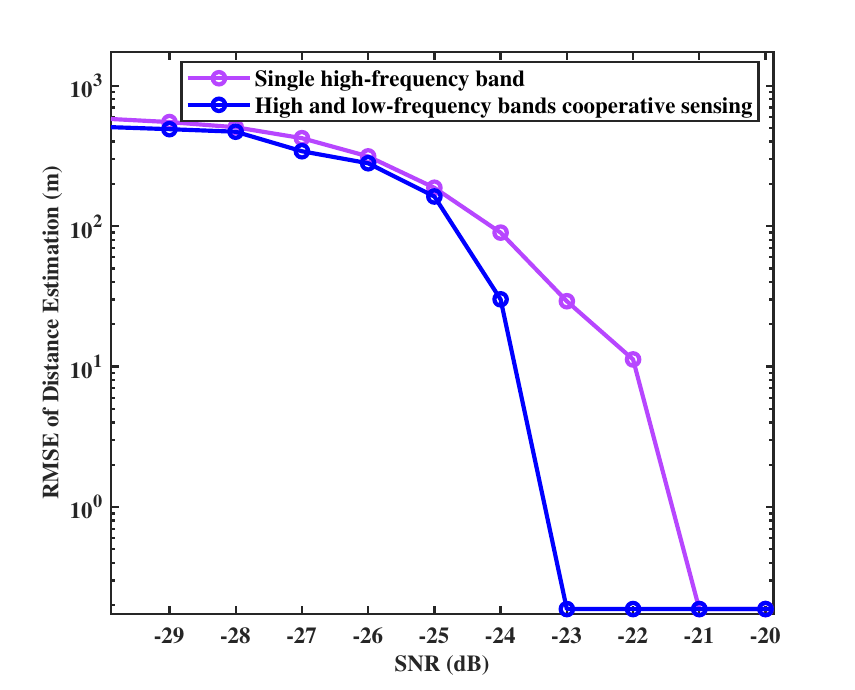}}
	\subfigure[Velocity estimation.] {\label{fig5.b}\includegraphics[width=0.4\textwidth]{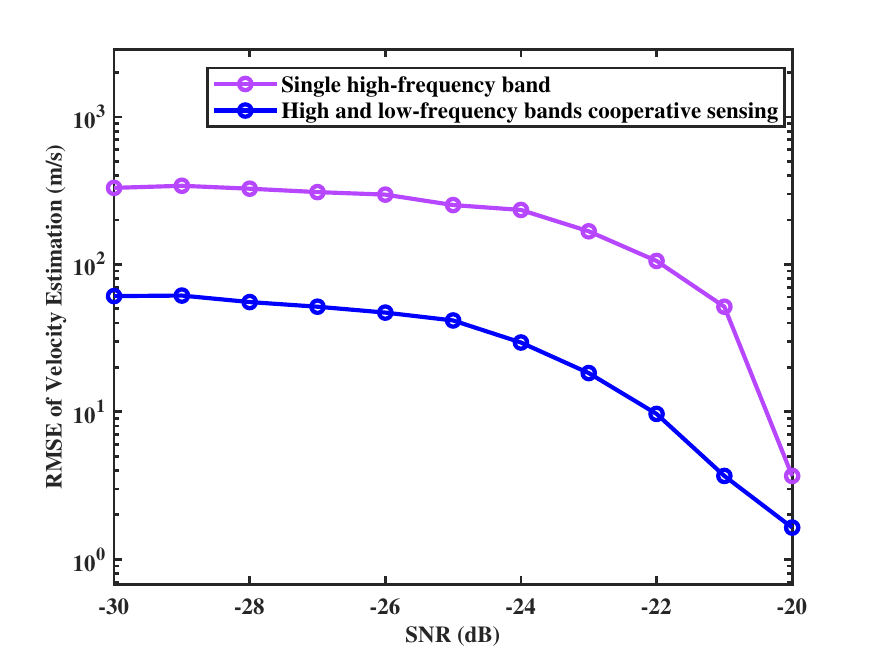}}
	\caption{RMSEs of multi-band sensing and single-band sensing.}
	\label{fig.5}
\end{figure}

According to the 3rd Generation Partnership Project (3GPP) 
38.211 standard,
the high-frequency band with carrier frequency of 24 GHz and subcarrier spacing of 120 kHz, 
and the low-frequency band with carrier frequency of 5.9 GHz and subcarrier spacing of 30 kHz, 
are applied in the high and low frequency bands cooperative sensing \cite{WeiCA}.
As illustrated in Fig. \ref{fig.5}, 
it is obvious that the RMSEs of distance and velocity estimation are 
lower and the convergence speed is faster 
with multiple frequency bands cooperative sensing 
than those results 
with single frequency band sensing.
The reason is that the time-frequency resources are extended 
with multiple frequency bands cooperation.

\begin{figure}[htbp]
\centering
\subfigure[Range estimation.]{\includegraphics[width=0.48\linewidth]{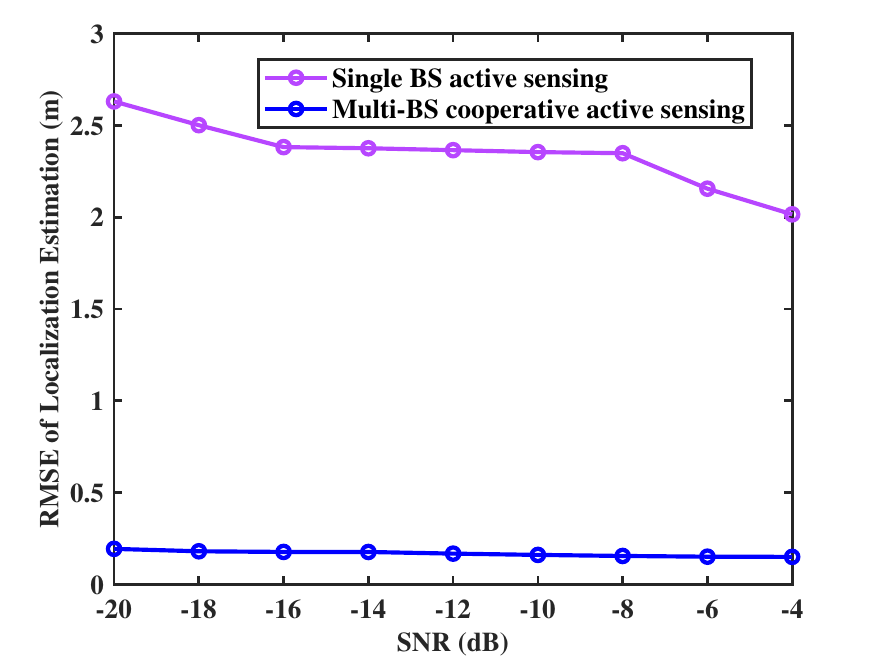}
\label{fig6(a)}}
\subfigure[Velocity estimation.]{\includegraphics[width=0.48\linewidth]{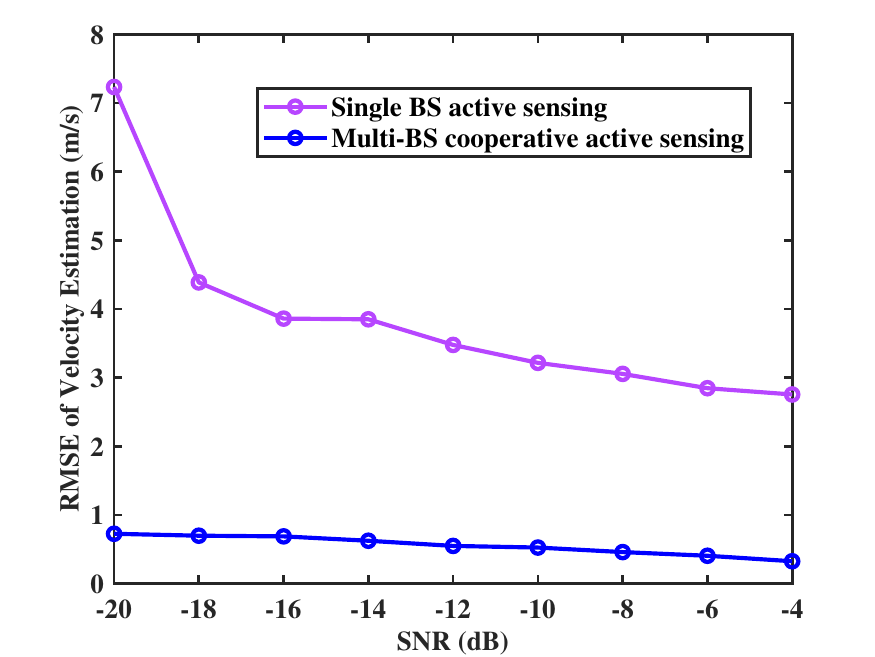}
\label{fig6(b)}}
\vspace*{0.5cm} % 这里是换行
\subfigure[ Multi-BS cooperative active-passive sensing.]{\includegraphics[width=0.48\linewidth]{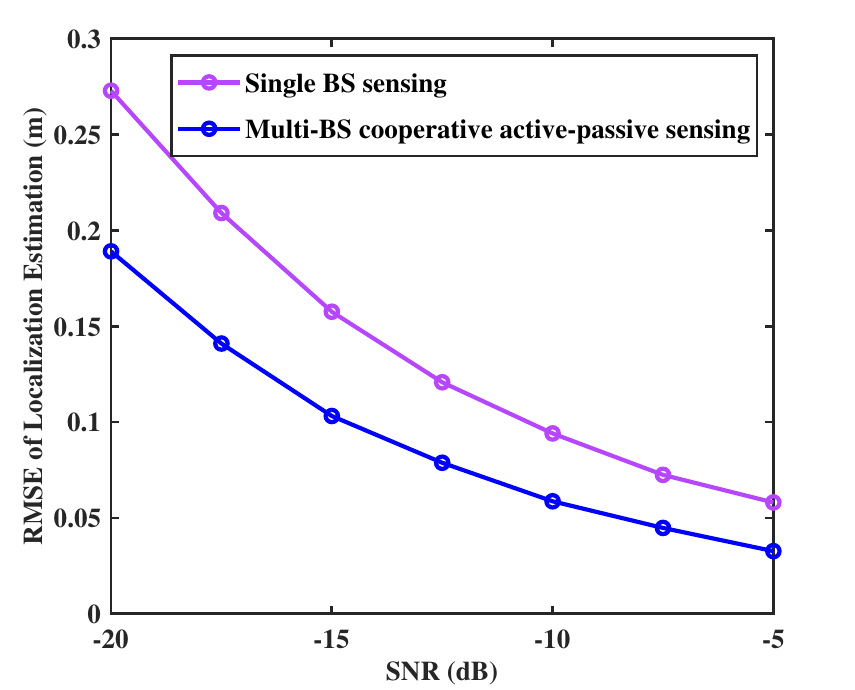}
\label{fig6(c)}}
\vspace*{0.5cm} % 这里是换行
\subfigure[Multi-BS cooperative passive sensing.]{\includegraphics[width=0.48\linewidth]{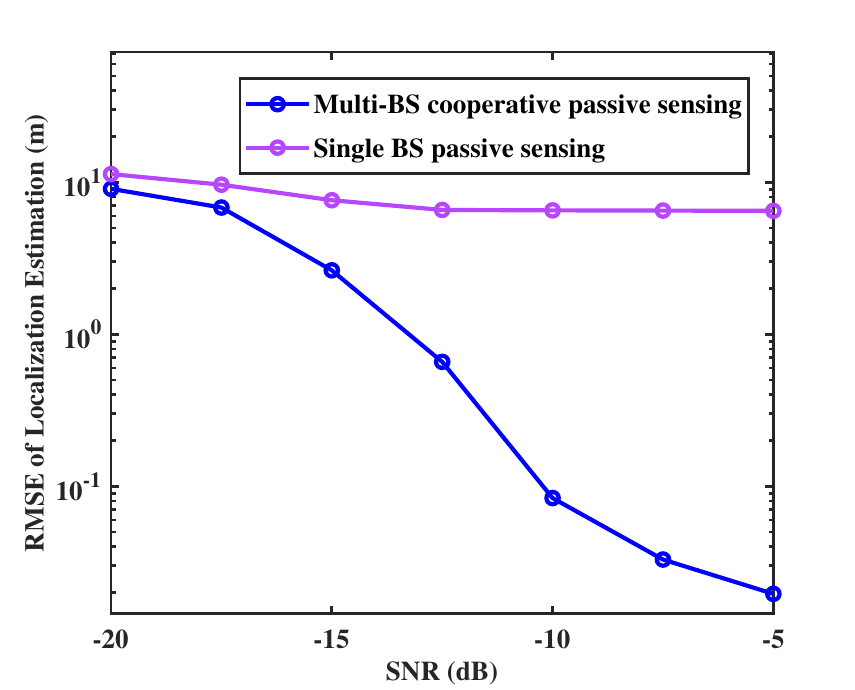}
\label{fig6(d)}}
\caption{RMSEs of multi-BSs cooperative sensing.} % 整体的标题
\label{fig6}
\end{figure}

\subsection{Node-level Cooperation}

As depicted in Fig. \ref{fig6}, the multi-BS 
cooperative UAV sensing is 
simulated to verify the performance of 
node-level cooperative sensing.
The number of BSs is 4, 
the subcarrier spacing is 120 kHz, and the carrier frequency is 4.9 GHz.
Multi-BS cooperative active sensing 
is compared with single-BS sensing in terms of 
the performance of localization and velocity estimation for UAV.
As illustrated in Figs. \ref{fig6(a)} and \ref{fig6(b)}, 
the multi-BS cooperative sensing achieves more 
accurate localization estimation
than single-BS sensing.
Multi-BS cooperative active sensing achieves significantly better
performance in velocity estimation than single-BS sensing.
The same simulation parameters are applied 
in multi-BS cooperative active-passive sensing and multi-BS cooperative passive sensing, where 
1000 times Monte Carlo simulations are performed to 
calculate the RMSE of location estimation.
It is revealed that the localization performance is significantly better with multi-BS cooperative active-passive sensing, as well as with multi-BS cooperative passive sensing, compared to single-BS sensing, as shown in Figs.~\ref{fig6(c)} and \ref{fig6(d)}.

\section{Conclusion}\label{se7}

In order to realize high-accuracy, 
large-scale, and continuous sensing in ISAC system,
this article provides a deep and comprehensive view on the cooperative sensing 
in ISAC system, including resource-level cooperative sensing,
node-level cooperative sensing, and
infrastructure-level cooperative sensing.
In the resource-level cooperation,
the sensing information in time-frequency-space-code domains 
is fused to improve sensing accuracy.
In node-level cooperation, multiple nodes, including BS and UE, could perform cooperative sensing
to fuse the sensing information from multiple nodes, 
extending the sensing coverage and improving 
the sensing accuracy.
In infrastructure-level cooperation,
the large number of infrastructures perform cooperative sensing 
to realize continuous sensing.
The research in this article may provide 
a research guideline for cooperative sensing in ISAC system,
promoting the applications of IoE with 
the connection of digital and physical spaces.

% reference
\bibliographystyle{IEEEtran}
\bibliography{reference}

\section*{Biography}
\textbf{Zhiqing Wei} (Member, IEEE) received the B.E. and Ph.D. degrees from the Beijing University of Posts and Telecommunications (BUPT), Beijing, China, in 2010 and 2015, respectively. He is an Associate Professor with BUPT. He has authored one book, three book chapters, and more than 50 papers. His research interest is the performance analysis and optimization of intelligent machine networks. He was granted the Exemplary Reviewer of IEEE WIRELESS COMMUNICATIONS LETTERS in 2017, the Best Paper Award of WCSP 2018. He was the Registration Co-Chair of IEEE/CIC ICCC 2018, the publication Co-Chair of IEEE/CIC ICCC 2019 and IEEE/CIC ICCC 2020.

\textbf{Haotian Liu} (Student Member, IEEE) received the B.E. degree in School of Physic and Electronic Information Engineering, Henan Polytechnic University (HPU) in 2023. He is currently pursuing his M.S. degree with Beijing University of Posts and Telecommunication (BUPT). His research interests include integrated sensing and communication, cooperative sensing, compressed sensing, carrier aggregation.

\textbf{Zhiyong Feng} (M'08-SM'15) received her B.E., M.E., and Ph.D. degrees from Beijing University of Posts and Telecommunications (BUPT), Beijing, China. She is a professor at BUPT, and the director of the Key Laboratory of the Universal Wireless Communications, Ministry of Education, P.R.China. She is a senior member of IEEE, vice chair of the Information and Communication Test Committee of the Chinese Institute of Communications (CIC). Currently, she is serving as Associate Editors-in-Chief for China Communications, and she is a technological advisor for international forum on NGMN. Her main research interests include wireless network architecture design and radio resource management in 5th generation mobile networks (5G), spectrum sensing and dynamic spectrum management in cognitive wireless networks, and universal signal detection and identification.

\textbf{Huici Wu} (Member, IEEE) received the Ph.D degree from Beijing University of Posts and Telecommunications (BUPT), Beijing, China, in 2018. From 2016 to 2017, she visited the Broadband Communications Research (BBCR) Group, University of Waterloo, Waterloo, ON, Canada. She is now an Associate Professor at BUPT. Her research interests are in the area of wireless communications and networks, with current emphasis on collaborative air-to-ground communication and wireless access security.

\textbf{Fan Liu} (Senior Member, IEEE) received the
B.Eng. and Ph.D. degrees from Beijing Institute of
Technology (BIT), Beijing, China, in 2013 and 2018
respectively, respectively. He is currently an Assistant Professor with the School of System Design
and Intelligent Manufacturing (SDIM), Southern
University of Science and Technology (SUSTech).
He has previously held academic positions with the
University College London (UCL), first as a Visiting
Researcher from 2016 to 2018, and then as a Marie
Curie Research Fellow from 2018 to 2020. His
research interests include signal processing and wireless communications, and in particular in the area of Integrated Sensing and Communications
(ISAC). 

\textbf{Qixun Zhang} (Member, IEEE) received the B.E.
and Ph.D. degrees from the Beijing University of
Posts and Telecommunications (BUPT), Beijing,
China, in 2006 and 2011, respectively. From March
2006 to June 2006, he was a Visiting Scholar with
the University of Maryland, College Park, MD,
USA. From November 2018 to November 2019,
he was a Visiting Scholar with the Department
of Electrical and Computer Engineering, University
of Houston, Houston, TX, USA. He is currently
a Professor with the Key Laboratory of Universal
Wireless Communications, Ministry of Education, School of Information and Communication Engineering, BUPT. His research interests include 5G mobile communication systems, integrated sensing and communication for autonomous driving vehicle, mmWave communication systems, and unmanned aerial vehicles (UAVs) communication.

\textbf{Yucong Du} (Student Member, IEEE) received the B.S. degree and the M.S. degree from Chongqing University of Posts and Telecommunications, Chongqing, China, in 2020 and 2023. He is currently pursuing the Ph.D. degree with the School of Information and Communication Engineering, Beijing University of Posts and Telecommunications (BUPT), Beijing, China. His current research interests include integrated sensing and communication network, digital twin network and resource management.

% \newpage

% \setcounter{figure}{0}

% \begin{figure*}
%     \centering
%     \includegraphics[width=0.99\linewidth]{fig1.pdf}
%     \caption{Resource-level cooperation in multi-domain resources.}
% \end{figure*}
% \clearpage

% \begin{figure*}
% 	\centering
%     \includegraphics[width=0.9\linewidth]{fig2.pdf}
%     \caption{Multi-BS cooperation.}
% \end{figure*}
% \clearpage

% \begin{figure*}
% 	\centering
%     \includegraphics[width=0.85\linewidth]{fig3.pdf}
%     \caption{BS-UE cooperation.}
% \end{figure*}
% \clearpage

% \begin{figure*}
% 	\centering
%     \includegraphics[width=0.85\linewidth]{fig4.pdf}
%     \caption{Infrastructure-level cooperation.}
% \end{figure*}
% \clearpage

% \begin{figure}
% 	\centering
% 	\subfigure[Range estimation.] {\includegraphics[width=0.4\textwidth]{fig5.a.pdf}}
% 	\subfigure[Velocity estimation.] {\includegraphics[width=0.4\textwidth]{fig5.b.pdf}}
% 	\caption{RMSEs of multi-band sensing and single-band sensing.}
% \end{figure}
% \clearpage

% \begin{figure}
% \centering
% \subfigure[Range estimation.]{\includegraphics[width=0.48\linewidth]{fig6.a.pdf}
% }
% \subfigure[Velocity estimation.]{\includegraphics[width=0.48\linewidth]{fig6.b.pdf}
% }
% \vspace*{0.5cm} % 这里是换行
% \subfigure[ Multi-BS cooperative active-passive sensing.]{\includegraphics[width=0.48\linewidth]{fig6.c.pdf}
% }
% \vspace*{0.5cm} % 这里是换行
% \subfigure[Multi-BS cooperative passive sensing.]{\includegraphics[width=0.48\linewidth]{fig6.d.pdf}
% }
% \caption{RMSEs of multi-BSs cooperative sensing.} % 整体的标题
% \end{figure}
% \clearpage

\end{document}